\documentstyle[prl,aps,epsf,multicol,psfig]{revtex}

\begin{document}

\title{ Facet ridge end points in crystal shapes}
\author{ Douglas Davidson and Marcel den Nijs}
\address{ Department of Physics, University of Washington, P.O. Box 351560, \\
Seattle, Washington 98195-1560}
\maketitle

\begin{abstract}

Equilibrium crystal shapes (ECS) near facet ridge end points (FRE) are
generically complex.
We study the body-centered solid-on-solid model on a square lattice
with an enhanced uniaxial interaction range
to  test the stability of the so-called stochastic FRE point
where the model maps exactly onto one dimensional
Kardar-Parisi-Zhang type
growth and the local ECS is simple.
The latter is unstable.
The generic ECS contains
first-order ridges extending into the rounded part of the ECS,
where two rough orientations coexist and 
first-order faceted to rough boundaries terminating in
Pokrovsky-Talapov type end points.

PACS number(s):  64.60 Fr, 68.35 Bs, 64.60 Ht, 68.35 Rh
\end{abstract}

\begin{multicols}{2}
\narrowtext

\begin{figure}
\centerline{\epsfxsize=5cm \epsfbox{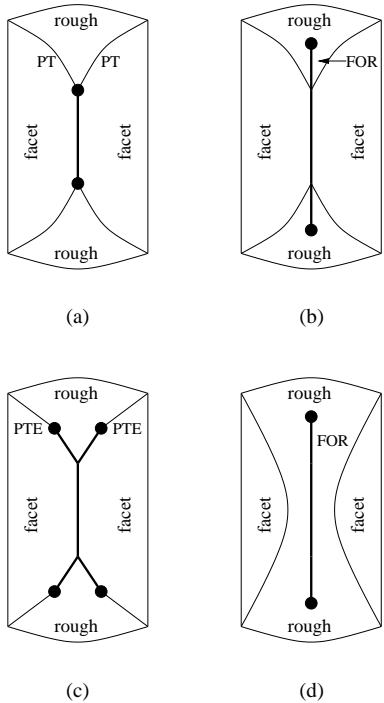}}
\vspace{0.1in}
\caption{Equilibrium crystal shapes in the BCSOS model with enhanced
interaction range:
(a) ECS in the exactly soluble square lattice BCSOS model with
    stochastic FRE point.
(b) ECS with a first-order line extending into the rough area.
(c) ECS with first-order facet-to-round boundaries and PTE points.
(d) ECS with a spontaneous tilted rough phase, i.e.,
    with a first-order ridge inside the rough phase.}
\label{ridge}
\end{figure}

Equilibrium crystal shapes (ECS) have been studied over the span of
many years ~\cite{review}.
Many features of their thermal evolution are theoretically well understood
and have been observed experimentally.
For example,
surface roughening of facets are realizations the
Kosterlitz-Thouless transition~\cite{review}.
They have been observed in Helium crystals ~\cite{Helium},
metal surfaces~\cite{metals}, ionic solids~\cite{ions}
and organic crystals~\cite{Bennema}.
Smooth boundaries between flat facets and rounded rough regions,
like in Fig.~\ref{ridge}(a), belong to the so-called
Pokrovsky-Talapov (PT) universality class.
They have been observed  in, e.g., lead~\cite{Metois} 
and Helium crystals~\cite{Helium2}.
An aspect for which experimental data is still sparse
is the ECS structure near facet ridge end point (FRE).
For example, in NaCl crystals, facet ridges seem to vanish very quickly
as temperature is increased~\cite{Metois2}.
Many other experiments reproduce growth shapes, where
FRE points are not represented properly
(rough surface orientations are suppressed 
because they grow faster than facets).

Lately the ECS structure  near FRE points has gained theoretical attention,
with the discovery of a link 
with Kardar-Parisi-Zhang (KPZ) type non-equilibrium
growth~\cite{Dennijs}.
Fig.~\ref{ridge}(a) represents the simplest and until now
canonical ECS structure near FRE points.
The sharp ridge between two faceted orientations terminates 
and splits into two PT facet-to-round boundaries.
This structure is realized in the so-called
body-centered solid-on-solid (BCSOS) model with next-nearest neighbor
interactions only~\cite{Bukman}, which is equivalent
to the exactly solved six-vertex model.
The FRE point where the PT lines join the facet ridge is very intriguing.
At this point, the model can be mapped
exactly onto a one dimensional BCSOS growth model that lies in the
KPZ universality class~\cite{Dennijs}.
To be more precise, the transfer matrix of the two dimensional
equilibrium model maps onto the master equation of the one
dimensional growth model. The spatial direction parallel to the facet
ridge plays the role of time in the dynamic process.

This seems to settle the issue, but the  FRE point in this exactly soluble
model
has too much symmetry (by being stochastic)
to leave us confident that its scaling behavior is universal and
that the ECS of Fig.~\ref{ridge}(a) represents the generic shape.
Recently, we performed a study of
the ECS in a BCSOS model on a honeycomb lattice~\cite{Davidson}.
There we did not find the structure of Fig.~\ref{ridge}(a), but
instead shapes equivalent  to  Fig.~\ref{ridge} (b) and (c).
In (b) the facet ridge extends into the rough region
as a first-order rough-to-rough (FOR) line.
It terminates at a critical end point.
This possibility had been suggested before~\cite{Bukman}
but had not been observed in any model beyond mean field theory.
Our result was somewhat unsatisfactory  however since
the FOR line remained extremely short.
In another part of the phase diagram we found an ECS similar to
Fig.~\ref{ridge}(c).
The boundaries between the faceted and the rounded rough regions are
partially first-order. These segments became quite long
and we could establish the scaling behaviour of the
critical end points (PTE points) where the transition becomes PT like.
PTE scaling  is in agreement with a simple theory where the free
energy is assumed analytic on the rough side
and expandable in powers of the tilt angle~\cite{Davidson,Shenoy,Miscut}.
PTE points have been observed in Si(113)~\cite{Song},
but with scaling exponents different from the above mean field type
values.
This can be attributed to long range attractive step-step
interactions in the context of step bunching~\cite{Miscut,Lassig}.

In this letter, we present a study of the BCSOS model on
the square lattice with enhanced interaction range.
We address directly the stability of the stochastic FRE point.
The FOR lines are well developed in this model.
They appear also as stand-alone ridges inside the rounded parts
of the crystal, as shown in Fig.~\ref{ridge}(d), 
representing spontaneous symmetry breaking within the
rough phase (coexistence between two rough surface orientations).

Consider a square lattice, oriented diagonally.
The most general Hamiltonian
with only next-nearest neighbor interactions, the one that is
exactly soluble,  can be written as
\begin{eqnarray}\label{Ham6v}
H_0 = \sum_{x,y} & [{1\over 4}L_H(h_{x+2,y}-h_{x,y}-E_H/L_H)^2 \nonumber\\
              +  & {1\over 4}L_V(h_{x,y+2}-h_{x,y}-E_V/L_V)^2 ]
\end{eqnarray}
with $x+2,y$ ($x,y+2$) and $x,y$ next nearest neighbors
in the horizontal (vertical) direction and
$h_{x,y}=0,\pm1,\pm2,\cdots$ the height variables.
Nearest neighbor heights must differ by one.
The fields $E_H$ and $E_V$ couple to the slopes, which we denote
as $Q_H$ and $Q_V$ respectively.  In this diagonal set up, the time
like direction of the transfer matrix coincides with the facet ridges
in Fig.~\ref{ridge}.
The ECS of this model takes the structure of Fig.~\ref{ridge}(a)
when $\exp(-L_H) > \exp(-L_V) +1$, in particular when 
the step-step interactions are attractive in the horizontal direction
and repulsive
in the vertical direction.
The FRE point has the stochastic KPZ character mentioned above
~\cite{Bukman,Dennijs}.
To study the stability of the stochastic FRE point
we need to break the special symmetries of $H_0$.
Increasing the interaction range will do this.
We choose uniaxial interactions for numerical convenience.
The simplest is to add interactions between next-to-next nearest neighbors
lying in the same horizontal row
\begin{eqnarray}\label{Ham}
H = H_0+  \sum_{x,y} [ & M_2\, \delta (h_{x,y},h_{x+4,y}\pm 2)  \nonumber \\
                     + & M_4\, \delta (h_{x,y},h_{x+4,y}\pm 4)]
\end{eqnarray}
with $\delta(x)$ Kronecker delta's.
This involves two coupling constants, $M_2$ and $M_4$,
because those further neighbors can differ in height by $0$, $\pm 2$, or
$\pm 4$.
The details of the numerical analysis are
the same as in our previous paper~\cite{Davidson} and outlined there in detail.
The eigenvalues of the transfer matrix
yield the exact free energy as function of tilt angle, $f(E_V,Q_H,N)$,
at all temperatures in a semi-infinite strip geometry of width $N$.
The ECS follows from this by a $N\to \infty$ finite size scaling analysis,
combined  with a Legendre transform.
The latter is in essence the Wulff construction~\cite{Andreev}.

\begin{figure}
\centerline{\epsfxsize=7.5cm \epsfbox{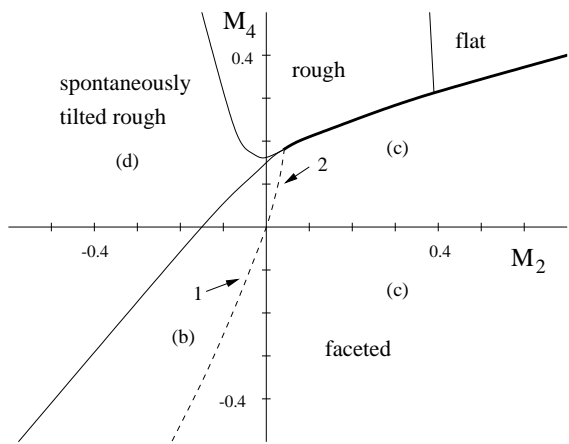}}
\vspace{0.05in}
\caption{Phase diagram of the model of Eq.~\ref{Ham} for $L_H=-0.69$,
$L_V=0.69$. The labels (b), (c), and (d) 
refer to the ECS in Fig.~\ref{ridge}, and $1$ and
$2$ to Fig.~\ref{split}(a) and (b).}
\label{phsdgr}
\end{figure}

The phase diagram at  $L_H=-0.69$ and $L_V=0.69$ for zero tilt fields,
$E_H=E_V=0$, is shown in Fig.~\ref{phsdgr}.
This is a representative example.
In the upper right corner,
where $M_2$ and $M_4$ are both large and positive, the surface is flat.
As $M_4$ is decreased, the surface
undergoes a first-order transition to a faceted phase.
The slope of the surface changes abruptly from $(0,0)$ to two
coexisting $(Q_H,Q_V)=(\pm 1,0)$ facets.
As $M_2$ is decreased, starting from the flat phase, the surface
first roughens, via a KT transition and 
then enters a spontaneously tilted rough phase~\cite{loc-note}.
This is a rough-to-rough faceting transition
with yet unknown anisotropic scaling properties.
In the tilted phase, two rough phases coexist.
The rounded part of the  ECS  contains a first-order ridge
as shown in Fig.~\ref{ridge}(d).
The jump in orientation
is caused by the competition between
the $L$ and $M$ interactions, and
changes continuously throughout this phase.
It locks in smoothly to $Q_H=\pm1$ at the PT boundary
into the faceted phase.

The dashed line in Fig.~\ref{phsdgr}
inside the faceted phase does not represent a phase transition.
It marks the boundary between the two different shapes,
Figs.~\ref{ridge} (b) and (c).
In region (c), the edge between each
facet and the rounded rough region is sharp near the FRE point.
The scaling at the PTE point (where the edge becomes a 
conventional PT boundary)
is consistent with our previous study,
i.e., with an analytic free 
energy functional in terms of the
surface tilt angle.
The correlation length in the direction along
the facet ridge scales compared to the perpendicular one as
$\xi_{\parallel}\sim l_{\perp}^z$ with $z=3$ ~\cite{Davidson,Shenoy,Miscut}.

In region (b) the first-order facet ridge boundary between the $(\pm 1,0)$
facets extends into the rounded region as a FOR line, as shown in
Fig.~\ref{ridge}(b).
Unlike our previous study on the honeycomb lattice~\cite{Davidson},
the FOR line becomes quite long and is unambiguously resolved.  For example,
at point $(L_H,L_V,M_2,M_4) =(-0.69,0.69,-0.02,0.02)$ the end point of
the FOR line lies at $E_V=1.04\pm0.01$, while the location of the
FRE point is known exactly, $E_V=0.908$.
At the phase boundary  into the spontaneously tilted rough phase, the ECS
changes from  Fig.~\ref{ridge}(c) to  Fig.~\ref{ridge}(d).
In the latter the facet-to-round boundary never touches the FOR line.

\begin{figure}
\centerline{\psfig{file=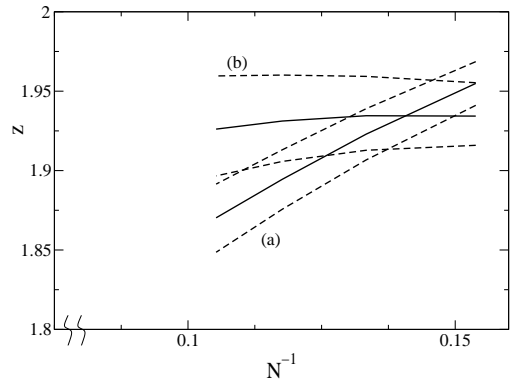,height=5cm,angle=270}}
\vspace{0.15in}
\caption{The effective dynamic exponent
as a function of strip width $N$ at the end points of two FOR line:
(a) $(L_H, L_V, M_2, M_4) = (-0.69, 0.69, -0.02, 0.02)$ and
(b) $(L_H, L_V, M_2, M_4) = (-0.41, 0.69, 0.05, -0.05)$.  The dashed
lines represent the uncertainty in the location of the end
points.}
\label{exp}
\end{figure}

The scaling behaviour at the FOR end point is an important issue.
A naive guess is to assume the free energy is analytic
in terms of the tilt angles.
It should then take the asymptotic form
$f(Q_H,E_V) = A (E_V-E_V^*)Q_H^2 + B Q_H^4$, such
that at the FOR critical end point the
free energy scales like $f\sim Q_H^4$.
The free energy gap between the $Q_H = 0$ and $Q_H = 1/L$
sectors would behave as $L[f(1/L)-f(0)] \sim L^{-3}$,
which translates into an anisotropic scaling exponent $z=3$.
The latter is inconsistent with our numerical data.
Fig.~\ref{exp} illustrates the  finite size scaling estimates for $z$ at
two FOR end points, located respectively at
$E_V=1.05$ for $(L_H,L_V,M_2,M_4) =(-0.69,0.69,-0.02,0.02)$,
and at
$E_V=0.62$ for $(L_H,L_V,M_2,M_4)=(-0.41,0.69,0.05,-0.05)$.
We measure the largest and second largest eigenvalues,
$\lambda_0$ and $\lambda_1$,
in the transfer matrix sector with periodic boundary conditions.
The gap  $m = \ln(\lambda_0/\lambda_1)$
scales with system size like $m \sim N^{-z}$.
Fig.~\ref{exp} illustrates that the
finite size scaling values of $z$ are rather sensitive to
the estimate of the location of the endpoint, and
that corrections to scaling vary strongly throughout the phase diagram,
but $z$ is clearly smaller than $2$.
Therefore, the above naive description is incorrect.

A second natural guess for the scaling properties of the end points
of FOR lines
is that they belong to the KPZ universality class.
The transfer matrix is not stochastic, but maybe some of
its properties survive.
The numerical data in Fig.~\ref{exp} do not exclude the KPZ value  $z=1.5$.
To test this possibility in more detail, we check
for a discontinuity in surface orientation, $\Delta Q_V$, at the
end point in the direction
along the FOR line.
$Q_V$ jumps at the stochastic FRE point in Fig.~\ref{ridge}(a).
This  represents the growth velocity in $1+1$ dimensional KPZ dynamics.
At the end points of the FOR lines, on the other hand,
this discontinuity vanishes 
as function of system size $N$,
like $\Delta Q_V\sim N^{-x}$ with $x=2.2\pm 0.1$ at all
points we checked.
The discontinuity in surface tilt along the FOR line in the
perpendicular
direction, $\Delta Q_H$, acts as order parameter.  
It vanishes continuously,
probably as a powerlaw, $\Delta Q_H \sim |E_V-E_V^*|^\beta$.
The discreteness in strip width $N$ allows only
commensurate rational values for $Q_H$.
This makes an accurate scaling analysis impossible.
However, we are confident that $0.3 < \beta < 0.5$.

In Figs.~\ref{ridge} (b) and (c) the FRE point has first-order
characteristics. The crossover region in between these structures, marked by
the dashed line in Fig.~\ref{phsdgr},
is the only place we can hope  to find
Fig.~\ref{ridge}(a) type structures.
However, this crossover region is actually blurred and contains
more complex crystal shapes.
At the stochastic points $M_2$=$M_4$=$0$ the crossover is simple.
On approach from the (b) side,
the FOR line shrinks in length and vanishes completely, followed by the
emergence of PTE points out of the FRE point on the (c) side.

\begin{figure}
\centerline{\epsfxsize=7cm \epsfbox{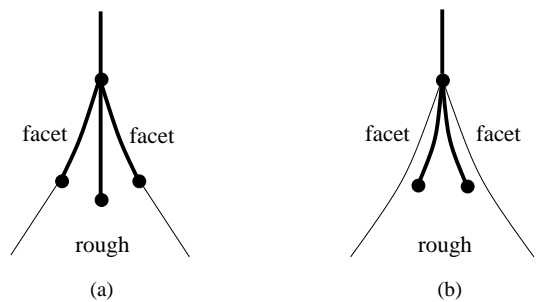}}
\vspace{0.1in}
\caption{The ECS near the FRE point in the crossover region
between the shapes of Fig.~\ref{ridge} (b) and (c).}
\label{split}
\end{figure}

\begin{figure}
\centerline{\psfig{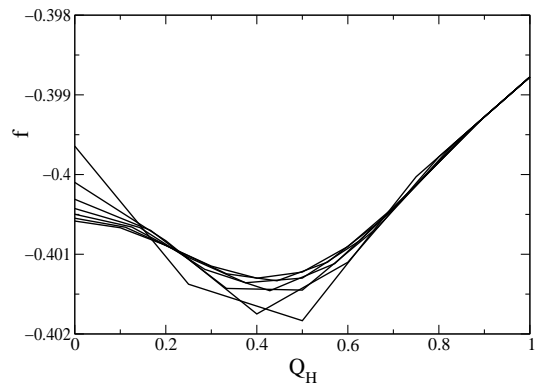}}
\vspace{0.15in}
\caption{The free energy as a function of $Q_H$ for 
$(L_H,L_V,M_2,M_4) =(0.0,0.69,-0.14,-0.40)$,
at $E_V=0.683$.  Only the $Q_H >0$ side is shown.  $f(Q_H)$ is mirror
symmetric.}
\label{fQH}
\end{figure}

On approach of the crossover region below the stochastic point,
marked 1 in Fig.~\ref{phsdgr}, the FOR line shrinks rapidly as well,
but just before it disappears completely, the
PTE points start to emerge already along
the two facet-to-round boundaries.
The intermediate ECS shape is a superposition of structures (b) and (c),
as shown in  Fig.~\ref{split}(a).
Fig.~\ref{fQH} shows the free energy as a function of surface tilt $Q_H$ at
$E_V=0.69$ for $(L_H,L_V,M_2,M_4)=(0,0.69,-0.14,-0.40)$.
The crucial feature is the simultaneous presence of
a central maximum at $Q_H=0$ followed by two nearby minima
(this creates the FOR line)
and concavity of the curve at the $Q_H=\pm 1$ edges
(the PTE points).
We refer to our other paper for a detailed discussion on how such features
translate into ECS properties~\cite{Davidson}.
The choppiness in Fig.~\ref{fQH} is due to the
limitation to rational values of $Q_H$ (multiples of $1/N)$.
The numbers are known to arbitrary accuracy and their
finite size scaling converges smoothly.

This superposition of the FOR and PTE shapes is predictable, because
it is hard to imagine how local features of $f(Q_H)$
at opposite tilt angles can  be topologically linked to each other,
except by special symmetries, like stochasticity.
The crossover region is very narrow, but numerically significant.
For example, in Fig.~\ref{fQH}
the FRE point lies at $(E_H,E_V) = (0,0.668 \pm 0.004)$ and
the FOR end points at $(0,0.70 \pm 0.01)$,
compared to the virtually exact location
$(\pm 0.15808,0.87105)$ of the PTE points.

Above the stochastic point, marked 2 in Fig.~\ref{phsdgr},
the crossover is more intricate.
Here, the free energy as a function of $Q_H$ develops extra wiggles and
a slightly concave part at intermediate $Q_H$.
On approach of the dashed line, the FOR line shortens dramatically
but then splits, see Fig.~\ref{split}(b).
The two FOR end points rapidly move outwards to merge
with the facet-to-round boundary into PTE points.

At the stochastic point the free energy curve $f(Q_H)$ is completely
featureless. It is flat. All tilt angles are degenerate,
due to the special symmetries associated with stochasticity.
In region 2 of the crossover region $f(Q_H)$ becomes almost flat,
but retains some wigglyness.
Some of this is thermodynamically unstable (skipped by the Legendre transform),
but the ECS is susceptible to complex hairy behaviour,
instead of simplifying into the featureless structure of Fig.~\ref{ridge}(a).
Complexity is indeed what we find,
although the splitting remains ambiguous, because
the splitted FOR lines are extremely short.
We can not rule out that their stability is a finite size
scaling artifact.
Therefore, we determined also the (effective) anisotropic exponent  $z$,
in region 2 pretending the ECS takes the simple form of Fig.~\ref{ridge}(a).
The results are consistent with the KPZ value $z=1.5$,
but this can be attributed to
crossover scaling from the stochastic point.

In conclusion,
the previous canonical picture of equilibrium crystal shapes
near facet ridge end points where the  facet ridge splits into two
PT lines, like in Fig.~\ref{ridge}(a), is too naive.
We established that the stochastic FRE point is unstable
and for the first time observed FOR lines unambiguously.
Generic experimental equilibrium crystal structures
near
FRE points will contain FOR lines sticking into the rough
rounded parts and first-order facet-to-round segments,
Figs.~\ref{ridge} (b) and (c) and Fig~\ref{split}.
We find also stand-alone FOR lines,
ridges inside the rounded part of the crystal where the
surface orientation jumps, as in Fig.~\ref{ridge}(d).
The scaling properties of the critical
end points of those ridges, and the critical point
at their birth (the rough to tilted phase boundary in Fig.~\ref{phsdgr})
need further study.
This research is supported by NSF grant DMR-9700430.

\end{multicols}

\begin{thebibliography}{30}
\bibitem{loc-note}
The precise location of these transitions is obscured by
modulations in $f(Q_H)$ that occur only in
this part of the phase diagram.  The transition
between the rough and tilted rough phases may be
partially first-order.
\end{thebibliography}
\end{document}